\documentclass[lettersize,journal]{IEEEtran}
\usepackage{amsmath,amsfonts}
\usepackage{algorithmic}
\usepackage{algorithm}
\usepackage{array}
\usepackage[caption=false,font=normalsize,labelfont=sf,textfont=sf]{subfig}
\usepackage{textcomp}
\usepackage{stfloats}
\usepackage{url}
\usepackage{verbatim}
\usepackage{multirow}
\usepackage{xcolor}
\usepackage{etoolbox}
\usepackage{graphicx}
\usepackage{cite}
\usepackage{threeparttable}
\usepackage{orcidlink}
\usepackage{lipsum}
\usepackage{fancyhdr}
\fancypagestyle{sec}{\lhead{\tmpx}}

\fancypagestyle{arXiv}
{
   \fancyhf{}
   \chead{\textcolor{gray}{This article was presented at the IEEE MTT-S International Microwave Symposium (IMS 2024), Washington, DC, USA, June 16–21, 2024 and invited to the IEEE Microwave and Wireless Technology Letters (MWTL).}}
   \cfoot{ \fontsize{=9pt}{10pt}\selectfont  \textcolor{gray}{© 2024 IEEE. Personal use of this material is permitted. Permission from IEEE must be obtained for all other uses, in any current or future media, including reprinting/republishing this material for advertising or promotional purposes, creating new collective works, for resale or redistribution to servers or lists, or reuse of any copyrighted component of this work in other works}\\\fontsize{=10pt}{12pt}\selectfont\thepage}
   
}

\makeatletter
\patchcmd{\@maketitle}
  {\addvspace{0.5\baselineskip}\egroup}
  {\addvspace{-1\baselineskip}\egroup}
  {}
  {}
\makeatother

\begin{document}

\title{MP-DPD: Low-Complexity Mixed-Precision Neural Networks for Energy-Efficient Digital Pre-distortion of Wideband Power Amplifiers}

\author{ Yizhuo~Wu\textsuperscript{*}\orcidlink{0009-0009-5087-7349},
    Ang~Li\textsuperscript{*}\orcidlink{0000-0003-3615-6755},
    Mohammadreza~Beikmirza\orcidlink{0000-0002-3886-0554},
    Gagan~Deep~Singh\orcidlink{0000-0002-0358-2451},
    Qinyu~Chen\orcidlink{0009-0005-9480-6164},\\
    Leo~C.~N.~de~Vreede\orcidlink{0000-0002-5834-5461},
    Morteza~Alavi\orcidlink{0000-0002-5834-5461},
    Chang~Gao\orcidlink{0000-0002-3284-4078}

\thanks{\textsuperscript{*}Equal Contributions.}
\thanks{Corresponding author: Chang Gao (chang.gao@tudelft.nl)}
\thanks{Yizhuo Wu, Ang Li, Mohammadreza Beikmirza, Gagan Deep Singh, Leo C. N. de Vreede, Morteza Alavi and Chang Gao are with the Department of Microelectronics, Delft University of Technology, The Netherlands.}
\thanks{Qinyu Chen is with the Leiden Institute of Advanced Computer Science (LIACS), Leiden University, The Netherlands.}
\thanks{This article was presented at the IEEE MTT-S International Microwave Symposium (IMS 2024), Washington, DC, USA, June 16–21, 2024}
}

\markboth{IEEE MICROWAVE AND WIRELESS TECHNOLOGY LETTERS}%
{Shell \MakeLowercase{\textit{et al.}}: A Sample Article Using IEEEtran.cls for IEEE Journals}


\maketitle

\begin{abstract}
Digital Pre-Distortion (DPD) enhances signal quality in wideband RF power amplifiers (PAs). As signal bandwidths expand in modern radio systems, DPD's energy consumption increasingly impacts overall system efficiency. Deep Neural Networks (DNNs) offer promising advancements in DPD, yet their high complexity hinders their practical deployment. This paper introduces open-source mixed-precision (MP) neural networks that employ quantized low-precision fixed-point parameters for energy-efficient DPD. This approach reduces computational complexity and memory footprint, thereby lowering power consumption without compromising linearization efficacy. Applied to a 160MHz-BW 1024-QAM OFDM signal from a digital RF PA, MP-DPD gives no performance loss against 32-bit floating-point precision DPDs, while achieving -43.75 (L)/-45.27 (R)\,dBc in Adjacent Channel Power Ratio (ACPR) and -38.72\,dB in Error Vector Magnitude (EVM). A 16-bit fixed-point-precision MP-DPD enables a 2.8$\times$ reduction in estimated inference power. The PyTorch learning and testing code is publicly available at \textcolor{red}{\url{https://github.com/lab-emi/OpenDPD}}.
\end{abstract}

\begin{IEEEkeywords}
digital pre-distortion (DPD), quantization, power amplifier (PA), deep neural network (DNN), digital transmitter (DTX)
\end{IEEEkeywords}

\section{Introduction}
\label{sec:introduction}
\thispagestyle{arXiv}
\IEEEPARstart{T}{he} rapid evolution of wireless communication technologies has spurred an increased demand for higher data rates, improved spectral efficiency, and reduced error rates. Non-linear distortions, predominantly caused by wideband Radio Frequency (\textbf{RF}) Power Amplifiers (\textbf{PAs}), significantly compromise signal integrity, affecting both communication reliability and energy efficiency. Digital Pre-Distortion (\textbf{DPD}) has emerged as a crucial technique to mitigate these issues, enhancing signal integrity. In contemporary radio digital front-ends, the DPD module is a major contributor to power consumption~\cite{wesemann2023energy}. This challenge might be further exacerbated by the potential integration of Machine Learning (\textbf{ML}) algorithms, such as Deep Neural Networks (\textbf{DNNs}), which, despite their potential, add to the power demands. 

Recent advancements of ML-based long-term DPD in state-of-the-art RF System-on-Chip (\textbf{SoC}) products are given in~\cite{ADRV9040}. Nevertheless, the substantial computational complexity and memory requirements of ML-based DPD systems, especially those using DNNs, pose significant obstacles to their efficient deployment in wideband transmitters, particularly in the context of future 5.5G/6G base stations or Wi-Fi 7 routers, where limited power resources constrain real-time DPD model computation.

Prior approaches to address DPD energy consumption include reducing the sample rate~\cite{Li2020SampleRate}, employing a sub-Nyquist feedback receiver in the observation path~\cite{Hammler2019}, dynamically adjusting model cross-terms based on input signal characteristics~\cite{Li2022}, and devising simpler computational pathways for DPD algorithms~\cite{Beikmirza2023}. This work presents a novel approach by implementing mixed-precision (\textbf{MP}) arithmetic operations and model parameters in a gated Recurrent Neural Network (\textbf{RNN})-based Digital Pre-distortion model for wideband PAs. The proposed method curtails the DPD model inference\footnote{Inference of a neural network model is the process of making predictions based on the learned model parameters. Learning in a model involves training the model to update the parameters with a dataset to classify patterns (classification) or to track a time-varying discrete variable (regression).} power consumption by substituting most high-precision floating-point operations with low-precision fixed-point operations through quantizing neural network weights (\textbf{W}) and activations (\textbf{A}). This strategy reduces the energy of arithmetic operations and memory access and facilitates the design of energy-and-area-efficient DNN computing hardware suitable for DPD deployment in power-sensitive environments~\cite{Gao2020}. Additionally, our method is compatible with existing strategies, allowing for further power savings when combined.
\section{The DPD Computing's Energy Problem}
To effectively correct the in-band signal and reduce out-of-band emission, DPD systems typically operate at sample rates ranging from 1.5$\times$ to 5$\times$ the baseband signal bandwidth~\cite{Li2020SampleRate}. As bandwidths in future radio systems expand, the energy demands of DPD computation intensify. The energy consumed per DPD model inference for each input I/Q sample is approximated by: 
\begin{equation}
    E_{\text{INF}} = E_{\text{MUL}} + E_{\text{ADD}} + E_{\text{MEM}}
    \label{eq:energy}
\end{equation}
where \( E_{\text{MUL}} \), \( E_{\text{ADD}} \), and \( E_{\text{MEM}} \) denote the energy consumption of multiplications (\textbf{MUL}), additions (\textbf{ADD}), and memory (\textbf{MEM}) access per DPD model inference, respectively. Since each inference processes one I/Q data point of the input signal, the estimated dynamic power consumption of DPD model inference is given as:
\begin{equation}
    P_{\text{INF}} = E_{\text{INF}} \cdot f_{s}
    \label{eq:power}
\end{equation}
where \( f_{s} \) represents the DPD input I/Q data sample rate.

\begin{figure}[t]
    \centering
    \includegraphics[width=\linewidth]{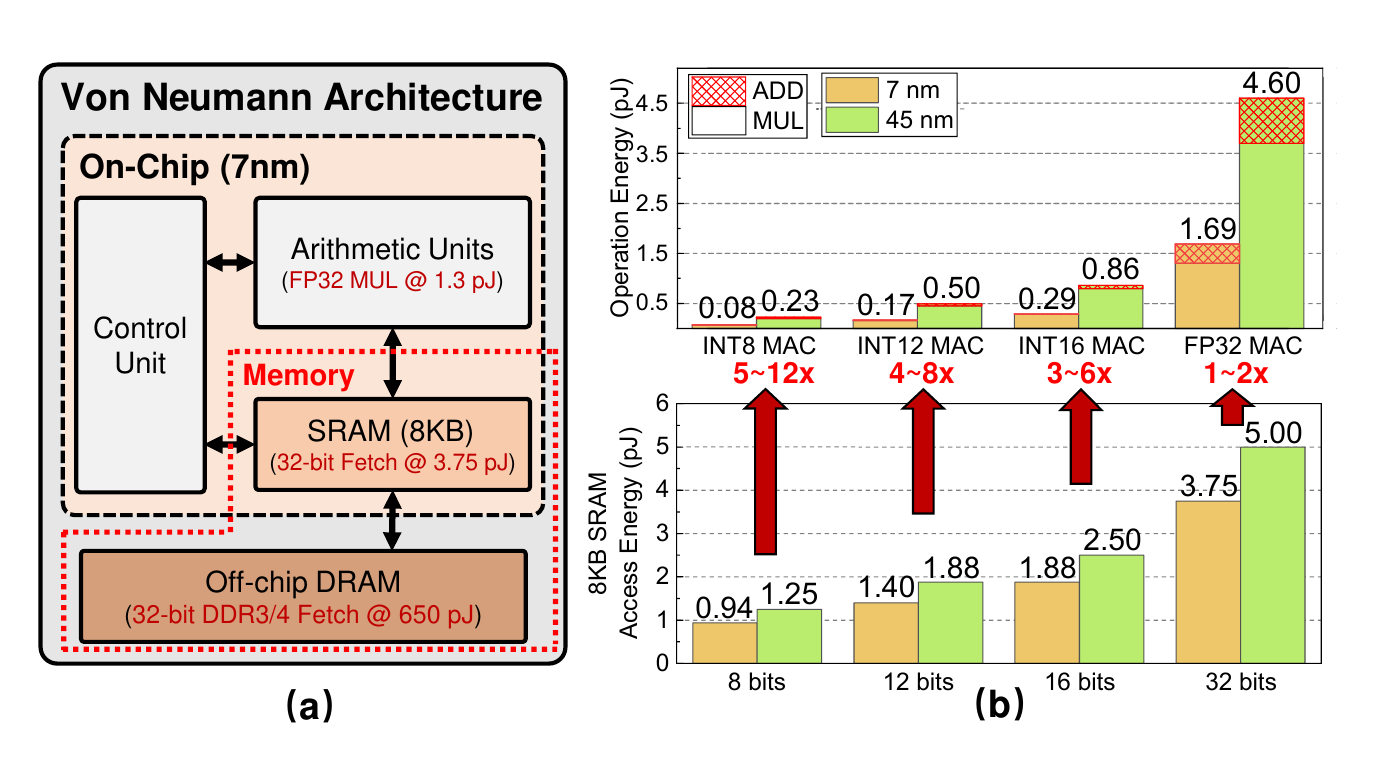}
    \caption{(a) The Von Neumann architecture with energy costs. (b) Operation and 8KB SRAM access energy in 45 nm \cite{horowitz20141} and 7 nm \cite{jouppi2021ten} vs. precision.}
    \label{fig:op_node_energy}
    \vspace{-12pt}
\end{figure}

Utilizing 32-bit floating-point (\textbf{FP32}) arithmetic, while beneficial for accuracy, can increase model size, negatively impacting energy efficiency. Prior studies demonstrate that DNNs with low-precision, fixed-point calculations effectively minimize the memory footprint in demanding applications such as image recognition and large language models. This reduction is achieved with minimal accuracy loss, decreasing power consumption in hardware implementations. As shown in Fig.~\ref{fig:op_node_energy}(b), Multiply-Accumulate (\textbf{MAC}) operations using 8-bit fixed-point integers (\textbf{INT8}) are up to 20$\times$ more energy-efficient than \textbf{FP32} MAC operations, across both 45nm~\cite{horowitz20141} and 7nm~\cite{jouppi2021ten} technology nodes. 
Most neural network computations occur on Von Neumann architecture-based hardware, depicted in Fig.~\ref{fig:op_node_energy}(a). This architecture often faces significant memory bottlenecks, as highlighted in Fig.~\ref{fig:op_node_energy}(b). The energy consumption of on-chip Static Random Access Memory (\textbf{SRAM}) is up to 12.2$\times$ higher than that of a MAC operation. Moreover, the energy costs for off-chip memory access are roughly three orders of magnitude greater than for arithmetic operations. Therefore, the memory access demands, directly linked to the DPD model size, play a crucial role in determining overall power consumption.

\section{Mixed-Precision Neural Networks DPD}
Building on these insights, this section describes how to quantize weights and activations of gated Recurrent Neural Networks (\textbf{RNNs}) into low precision for energy reduction.


\subsection{Gated Recurrent Unit-based DPD}
Gated RNNs utilize gates to manage information flow through their high-dimensional hidden states according to new input stimuli. This approach effectively addresses the vanishing gradient issue in modeling long sequences and makes them widely adopted in prior research on long-term DPDs~\cite{VDLSTM,DVRJANET}. In this work, the GRU-based DPD is defined as:
\begin{align}
\mathbf{r}_{t} &= \sigma\left(\mathbf{W}_{ir}\boldsymbol{\phi}_{t} + \mathbf{b}_{ir} + \mathbf{W}_{hr}\mathbf{h}_{t-1} + \mathbf{b}_{hr}\right) \label{eq:gru_0} \\
\mathbf{z}_{t} &= \sigma\left(\mathbf{W}_{iu}\boldsymbol{\phi}_{t} + \mathbf{b}_{iz} + \mathbf{W}_{hz}\mathbf{h}_{t-1} + \mathbf{b}_{hz}\right) \label{eq:gru_1} \\
\mathbf{n}_{t} &= \textrm{tanh}\left(\mathbf{W}_{in}\boldsymbol{\phi}_{t} + \mathbf{b}_{in} + \mathbf{r}_{t} \odot \left(\mathbf{W}_{hn}\mathbf{h}_{t-1} + \mathbf{b}_{hn}\right)\right) \label{eq:gru_2} \\
\mathbf{h}_{t} &= \left(1 - \mathbf{z}_{t}\right) \odot \mathbf{n}_{t} + \mathbf{z}_{t} \odot \mathbf{h}_{t-1} \label{eq:gru_3}
\end{align}
\noindent where \( \boldsymbol{\phi}_{t} \) is the input feature vector extracted from the I/Q modulated signal 
\(\mathbf{X} = \{\mathbf{x}_{t} | \mathbf{x}_{t} = I_\mathbf{x,t} + jQ_\mathbf{x,t}, I_\mathbf{x,t}, Q_\mathbf{x,t} \in \mathbb{R}, t \in 0,\dots ,T-1\}\) at time \( t \). \( \mathbf{h}_{t} \) represents the hidden state at time \( t \). The $\mathbf{W}$ and $\mathbf{b}$ terms are the weight matrices and bias vectors, respectively. The terms \( \mathbf{r}_{t} \), \( \mathbf{z}_{t} \), and \( \mathbf{n}_{t} \) correspond to the reset gate, update gate, and new candidate state, respectively. \( \sigma \) represents the sigmoid activation. $\odot$ denotes the element-wise multiplication. The GRU is followed by a fully-connected (\textbf{FC}) layer to generate the DPD output I/Q signal:
\begin{align}
\mathbf{\hat{y}}_{t} &= \mathbf{W}_{\hat{y}}\boldsymbol{h}_{t} + \mathbf{b}_{\hat{y}}
\label{eq:gru_4}
\end{align}
\noindent where \(\mathbf{\hat{y}}_{t} \in \mathbf{\hat{Y}} = \{\mathbf{y}_{t} | \mathbf{y}_{t} = I_{\mathbf{\hat{y}},t} + jQ_{\mathbf{\hat{y}},t}, I_{\mathbf{\hat{y}},t}, Q_{\mathbf{\hat{y}},t} \in \mathbb{R}, t \in 0,\dots ,T-1\}\).

\subsection{Mixed-Precision DPD}
\label{mixed-precision-gru}
To enhance the energy efficiency of DPD models, we adopt a mixed-precision strategy utilizing low-precision fixed-point integer arithmetic for inference. This method involves a quantization scheme that converts the model's weights and activations, including other intermediate variables, to lower precision while retaining full-precision operations for feature extraction $\boldsymbol{\phi}$ from I/Q signal $\mathbf{x}$, effectively balancing accuracy and computational complexity.

The quantization process is defined as follows: for a data point $x$, a quantization scale $s$, and a range $[Q_{\text{min}}, Q_{\text{max}}]$, the fixed-point representation $q$ of $x$ is calculated using:
\begin{equation}
q = s \times \texttt{Round}\left(\texttt{Clip}\left(\frac{x}{s}, Q_{\texttt{min}}, Q_{\texttt{max}}\right)\right) \label{eq_quant}
\end{equation}
where \texttt{Clip} bounds the input and \texttt{Round} rounds to the nearest integer. For $n$-bit quantization, unsigned data ranges from $Q_{\texttt{min}} = 0$ to $Q_{\texttt{max}} = 2^n - 1$, and signed data from $Q_{\texttt{min}} = -2^{n-1}$ to $Q_{\texttt{max}} = 2^{n-1} - 1$. During training, each neural network layer's quantization scale $s$ is optimized using gradient descent and adjusted to the nearest power-of-two, ensuring a fixed-point representation $q$.
\begin{figure}[t]
    \centering
    \includegraphics[width=1.0\linewidth]{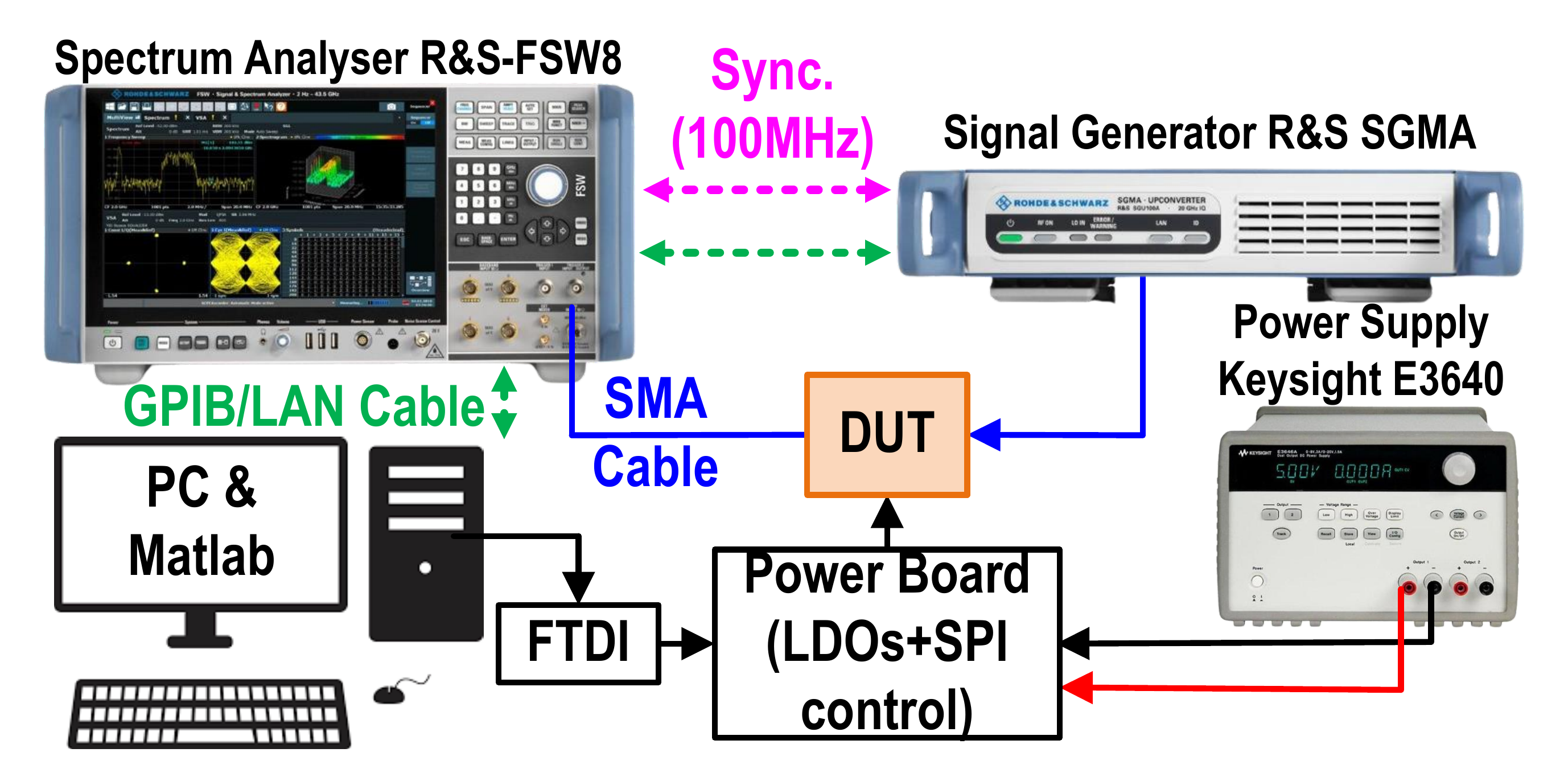}
    \caption{Setup for dataset acquisition and DPD performance measurement.}
    \label{fig:platform}
    \vspace{-12pt}
\end{figure}
For precise fixed-point computations and enhanced energy efficiency, we use a quantization-aware training method~\cite{Nagel2021}. This approach maintains full-precision variable copies updated during gradient descent while using quantized values for forward propagation of the DNN model. The gradient of the $\texttt{Round}$ function is approximated using the straight-through estimator \cite{bengio2013estimating} for trainability.

\begin{table*}[t]
\centering
\caption{ACPR and EVM Performance of Different DPD Models Evaluated with 160-MHz 4-Channel$\times$40\,MHz 1024-QAM OFDM Signals Sampled at 640\,MHz alongside Their Estimated Inference Energy and Dynamic Power Consumption in 7\,nm and 45\,nm~\cite{jouppi2021ten}.}
\label{tab:GRU}
\begin{threeparttable}
\begin{tabular}{|c|c|c|c|c|cc|cc|c|}
\hline
 &  &  &  &  & \multicolumn{2}{c|}{\textbf{Energy/Inference (nJ)}} & \multicolumn{2}{c|}{\textbf{Dynamic Power (W)}} &  \\ \cline{6-9}
\multirow{-2}{*}{\textbf{Classes}} & \multirow{-2}{*}{\textbf{\begin{tabular}[c]{@{}c@{}}DPD\\ Models\tnote{a}\end{tabular}}} & \multirow{-2}{*}{\textbf{\begin{tabular}[c]{@{}c@{}}ACPR \\ (dBc, L/R)\end{tabular}}} & \multirow{-2}{*}{\textbf{\begin{tabular}[c]{@{}c@{}}EVM\\ (dB)\end{tabular}}} & \multirow{-2}{*}{\textbf{\begin{tabular}[c]{@{}c@{}}Number of\\ MUL/ADD/MEM\end{tabular}}} & \multicolumn{1}{c|}{\textbf{45nm}} & \textbf{7nm} & \multicolumn{1}{c|}{\textbf{45nm}} & \textbf{7nm} & \multirow{-2}{*}{\textbf{\begin{tabular}[c]{@{}c@{}}Power\\ Reduction\end{tabular}}} \\ \hline
Without DPD & - & -31.69/-32.45 & -27.05 & - & \multicolumn{1}{c|}{-} & - & \multicolumn{1}{c|}{-} & - & - \\ \hline \hline
 & GMP~\cite{GMP} & -40.79/-40.86 & -29.27 & 2190/3668/517 & \multicolumn{1}{c|}{11.44} & 6.20 & \multicolumn{1}{c|}{7.32} & 3.97 & - \\ \cline{2-10} 
 & VDLSTM~\cite{VDLSTM} & -43.38/-43.02 & -36.19 & 538/1528/542 & \multicolumn{1}{c|}{3.38} & 3.32 & \multicolumn{1}{c|}{2.16} & 2.12 & - \\ \cline{2-10} 
 & RVTDCNN~\cite{Hu2022RVTDCNN} & -44.27/-43.50 & -36.70 & 500/2690/512 & \multicolumn{1}{c|}{4.28} & 3.60 & \multicolumn{1}{c|}{2.74} & 2.30 & - \\ \cline{2-10} 
\multirow{-4}{*}{FP32-DPDs} & \textbf{GRU} & \textbf{-43.36/-45.30} & \textbf{-38.46} & \textbf{502/1417/506} & \multicolumn{1}{c|}{\textbf{5.66}} & \textbf{3.09} & \multicolumn{1}{c|}{\textbf{3.62}} & \textbf{1.98} & \textbf{1$\times$} \\ \hline \hline
{\color[HTML]{000000} } & {\color[HTML]{FE0000} \textbf{W16A16-GRU}} & {\color[HTML]{FE0000} \textbf{-43.75/-45.27}} & {\color[HTML]{FE0000} \textbf{-38.72}} & {\color[HTML]{FE0000} \textbf{502/1417/506}} & \multicolumn{1}{c|}{{\color[HTML]{FE0000} \textbf{4.02}}} & {\color[HTML]{FE0000} \textbf{1.11}} & \multicolumn{1}{c|}{{\color[HTML]{FE0000} \textbf{1.93}}} & {\color[HTML]{FE0000} \textbf{0.71}} & {\color[HTML]{FE0000} \textbf{2.8$\times$}} \\ \cline{2-10} 
{\color[HTML]{000000} } & {\color[HTML]{FE0000} W12A16-GRU} & {\color[HTML]{FE0000} -43.03/-44.69} & {\color[HTML]{FE0000} -37.47} & {\color[HTML]{FE0000} 502/1417/506} & \multicolumn{1}{c|}{{\color[HTML]{FE0000} 2.29}} & {\color[HTML]{FE0000} 0.85} & \multicolumn{1}{c|}{{\color[HTML]{FE0000} 1.46}} & {\color[HTML]{FE0000} 0.54} & {\color[HTML]{FE0000} 3.7$\times$} \\ \cline{2-10} 
{\color[HTML]{000000} } & W12A12-GRU & -42.36/-43.79 & -37.45 & 502/1417/506 & \multicolumn{1}{c|}{2.19} & 0.82 & \multicolumn{1}{c|}{1.40} & 0.52 & 3.8$\times$ \\ \cline{2-10} 
{\color[HTML]{000000} } & W8A16-GRU & -41.64/-42.80 & -36.24 & 502/1417/506 & \multicolumn{1}{c|}{1.56} & 0.74 & \multicolumn{1}{c|}{1.00} & 0.47 & 4.2$\times$ \\ \cline{2-10} 
{\color[HTML]{000000} } & W8A12-GRU & -41.78/-42.90 & -36.17 & 502/967/506 & \multicolumn{1}{c|}{1.49} & 0.72 & \multicolumn{1}{c|}{0.95} & 0.46 & 4.3$\times$ \\ \cline{2-10} 
\multirow{-6}{*}{{\color[HTML]{000000} \textbf{\begin{tabular}[c]{@{}c@{}}MP-DPDs\tnote{b}\\ (This work)\end{tabular}}}} & W8A8-GRU & -35.84/-35.70 & -28.89 & 502/967/506 & \multicolumn{1}{c|}{1.42} & 0.69 & \multicolumn{1}{c|}{0.90} & 0.44 & 4.5$\times$ \\ \hline
\end{tabular}
\begin{tablenotes}
\item[a] The numbers of parameters are 495 (GMP), 502 (GRU), 538 (VDLSTM), 500 (RVTDCNN).
\item[b] Each MP-DPD has 14 and 17 FP32 MULs and ADDs for feature extraction, respectively.
\end{tablenotes}
\end{threeparttable}
\end{table*}

\section{Experimental Results}
\subsection{Experimental Setup}
Figure~\ref{fig:platform} illustrates the experimental setup. The baseband I/Q data was processed by a 40\,nm CMOS digital PA (\textbf{DPA})~\cite{Beikmirza2023jssc} at a 2.4\,GHz carrier frequency.

For the GRU-based MP-DPDs, quantization of activations and weights is performed at 8, 12, or 16 bits, except during feature extraction, which utilizes full-precision (FP32) operations to generate $I_x, Q_x, |x|, |x|^3$ features. We compared the MP-DPDs' performance to FP32 models, including General Memory Polynomial (\textbf{GMP})~\cite{GMP}, GRU, Vector Decomposition LSTM (\textbf{VDLSTM}), and Real-Valued Time-Delay Convolution Neural Network (\textbf{RVTDCNN}). The configurations for VDLSTM and RVTDCNN followed their optimal settings in~\cite{VDLSTM, Hu2022RVTDCNN}, with adjustments in model size through the hidden LSTM and FC layer sizes.

The test signal's Peak-to-Average Power Ratio (\textbf{PAPR}) is 10.38\,dB, and the DPA outputs at $13.75$\,dBm. The dataset, comprising 491,520 samples of 160-MHz 4-Channel$\times$40\,MHz OFDM signals sampled at 640\,MHz, was split into a 60\% training set for DPD learning, a 20\% validation set for early stopping, and a 20\% test set for performance evaluation. The DPD learning process involves backpropagation through a pre-trained PA model using our \texttt{OpenDPD}~\cite{wu2024opendpd} with the collected dataset in an end-to-end approach. All PA models consist of approximately 500 parameters, except for those used in parameter scan experiments. For both PA modeling and DPD learning, the models are trained for 100 epochs using the ADAM optimizer with a learning rate of 1E-3 and a batch size of 3200 samples.
\subsection{Results and Discussion}
\begin{figure}[t]
\centering\includegraphics[width=1.0\linewidth]{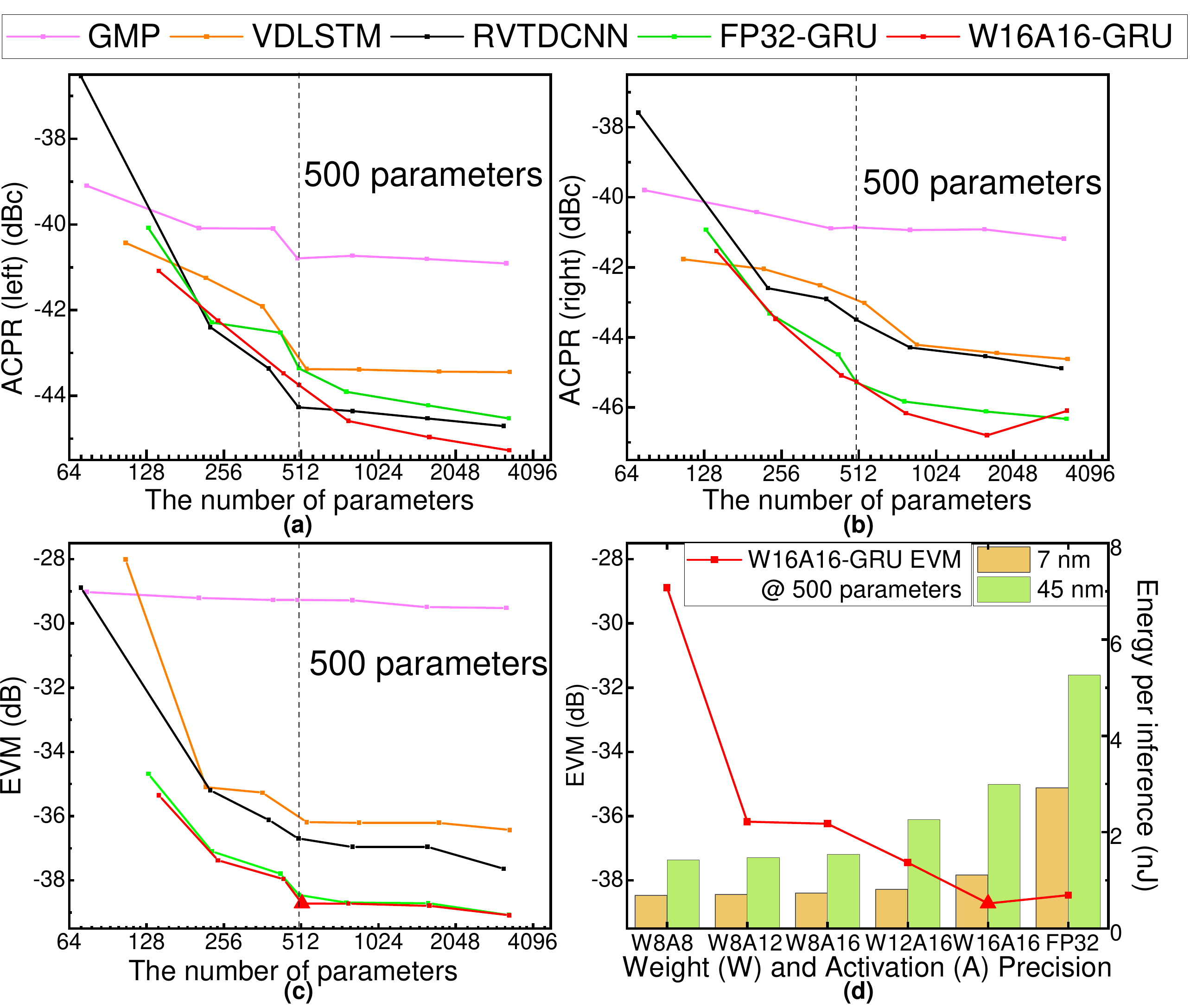}
    \caption{Parameter scan of DPD models vs. (a) ACPR (left) (b) ACPR (right) (c) EVM; (d) EVM (left Y-axis) and energy per inference (right Y-axis) vs. precision.}
    \label{fig:results}
\end{figure}
\begin{figure}[t]
    \centering
    \includegraphics[width=1.0\linewidth]{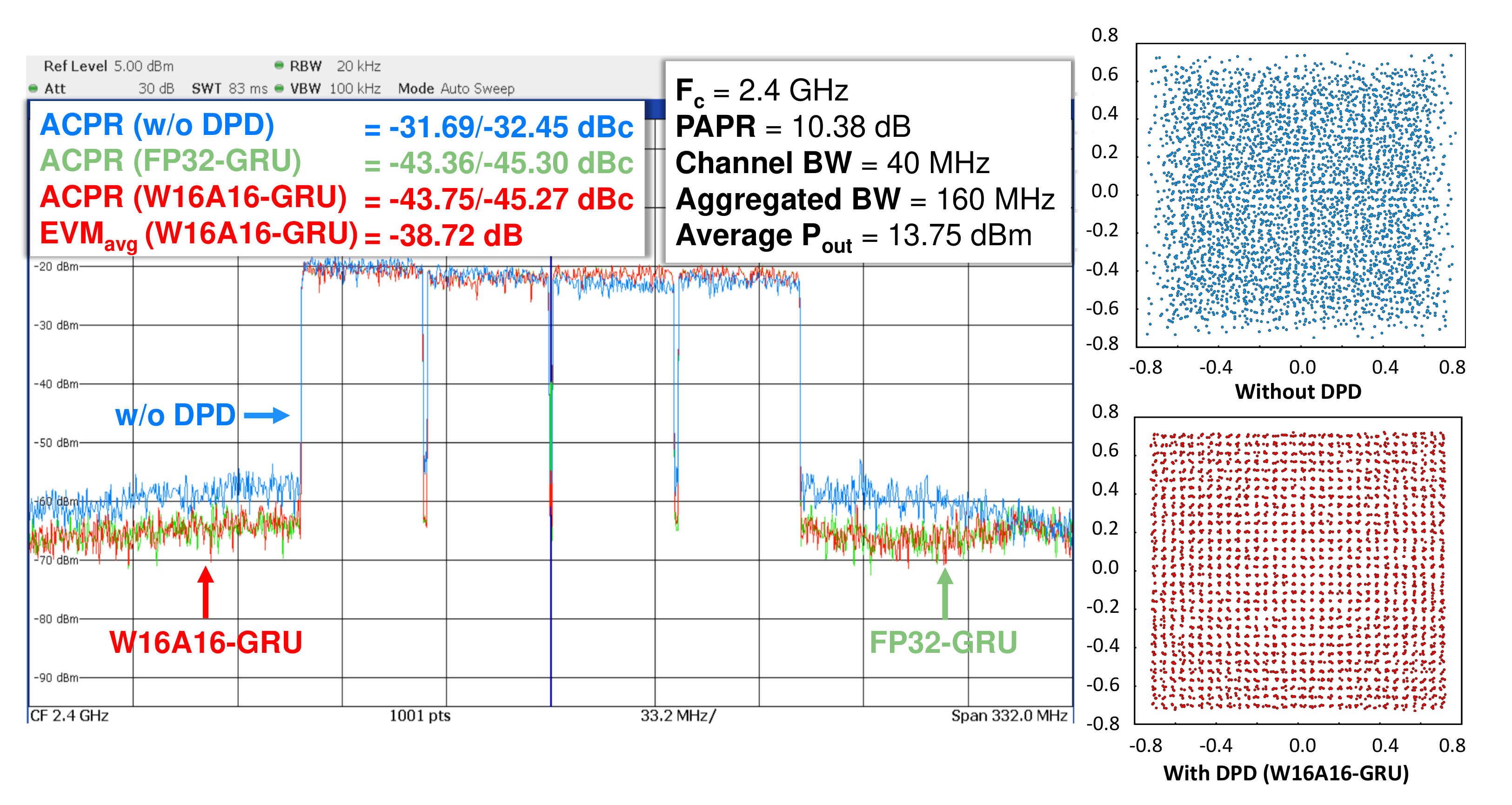}
    \caption{Measured spectrum and constellation map on the 160\,MHz signal.}
    \label{fig:PSD}
    \vspace{-12pt}
\end{figure}
Table~\ref{tab:GRU} compares the ACPR and EVM results for different DPD models, alongside the number of MUL, ADD operations, and 8KB SRAM accesses\footnote{Each input I/Q sample necessitates 2 input fetches, \#parameter fetches, and 2 output write-backs between the arithmetic units and the 8KB SRAM cache. Intermediate results are buffered locally, thus bypassing cache access.} in feature extraction and model inference (Eqs.~\eqref{eq:gru_0} $\sim$ \eqref{eq:gru_4}). The amplitude/phase (\texttt{arctan2}) group, \texttt{tanh}, and \texttt{sigmoid} functions can be computed using the COordinate Rotation DIgital Computer (\textbf{CORDIC}) algorithm over 15 iterations (30 ADDs) despite that state-of-the-art gated RNN hardware~\cite{Gao2022Spartus} uses look-up tables to approximate them with less energy and chip area overhead. The 502-parameter W16A16-GRU DPD model demonstrates the best performance among all tested models, achieving an ACPR of -43.36/-45.30\,dBc and an EVM of -38.72\,dB while consuming 1.13\,nJ per inference in 7nm technology and 0.72\,W dynamic power at 640 MHz. Lower power can be achieved by using a smaller model size or lower precision at the price of worse accuracy, as shown in Fig.~\ref{fig:results}.

Figs.~\ref{fig:results}(a)-(c) show the correlation between model size and ACPR/EVM, covering 100 to 3200 parameters. The W16A16-GRU DPD model notably outperforms FP32 models in many settings due to the regularization effect by training with quantization noise~\cite{Nagel2021}. Fig.~\ref{fig:results}(d) presents the energy efficiency versus performance trade-offs in MP models. The W8A8 model achieves a 4.5$\times$ power reduction over the FP32 model in 7nm technology at the expense of linearization performance. The W12A16 and W16A16 configurations present a balanced compromise, offering 3.7$\times$ and 2.8$\times$ less power consumption than the FP32 GRU baseline DPD model while sustaining competitive EVM. Hence, W12A16 and W16A16 are optimal for power-critical applications demanding high accuracy.

Fig.~\ref{fig:PSD} displays the measured spectrum and constellation map with and without DPDs. The spectrum analysis confirms that the W16A16-GRU model achieves no ACPR performance loss compared to the FP32-GRU model. 

These findings underscore the effectiveness of our MP-DPD approach in reducing DPD power consumption while sustaining linearization performance.
\subsection{Power Consumption Comparison to Prior Works}
Prior hardware implementations of DPD hardly reported any power consumption numbers~\cite{Declan2018,Christophe2013}. To our best knowledge, the only work we found is a sub-sampling DPD Field-Programmable Gate Array (\textbf{FPGA}) implementation~\cite{Li2022}, which consumes 1.875\,W to linearize 100\,MHz signal with a 150\,MHz sampling rate and 320 parameters. For a fair comparison, we normalized it to the sample rate we used in this paper, which is 640MHz. By adopting our proposed mixed-precision A16W16-GRU DPD with 502 parameters, the power consumption can be reduced by 3.9$\times$/10.6$\times$ to 1.93\,W/0.71\,W on a 45\,nm/7\,nm process, respectively.

\section{Conclusion}
This work proposes the MP-DPD method for wideband RF power amplifiers using the \texttt{OpenDPD} framework~\cite{wu2024opendpd}. This approach reduces the computational complexity against the full-precision baseline, thereby contributing to power savings while preserving superior linearization performance for more sustainable and energy-efficient wireless communication. 

\newpage
\bibliographystyle{IEEEtran}

\bibliography{IEEEabrv,ref.bib} 

\end{document}